\documentstyle[twocolumn,psfig]{article}
\makeatletter
\def\@normalsize{\@setsize\normalsize{12pt}\xpt\@xpt
\abovedisplayskip 10pt plus2pt minus5pt\belowdisplayskip \abovedisplayskip
\abovedisplayshortskip \z@ plus3pt\belowdisplayshortskip 6pt plus3pt
minus3pt\let\@listi\@listI}

\def\subsize{\@setsize\subsize{12pt}\xipt\@xipt}

\def\section{\@startsection {section}{1}{\z@}{24pt plus 2pt minus 2pt}
{12pt plus 2pt minus 2pt}{\large\bf}}

\def\subsection{\@startsection {subsection}{2}{\z@}{12pt plus 2pt minus 2pt}
{12pt plus 2pt minus 2pt}{\subsize\bf}}
\makeatother

\begin{document}

%don't want date printed
%\date{}
\date{To be published in: Proceedings of the Workshop on Physics and
Computation, PhysComp '94 (Los Alamitos: IEEE Comp. Soc. Press, 1994)}

%make title bold and 14 pt font (Latex default is non-bold, 16 pt)
\title{\Large\bf Results on two-bit gate design for quantum computers}

%for single author (just remove % characters)
%\author{I. M. Author \\
%  My Department \\
%  My Institute \\
%  My City, ST, zip}

%for two authors (this is what is printed)
\author{\begin{tabular}[t]{c@{\extracolsep{8em}}c}
  David P. DiVincenzo & John Smolin\cite{byline} \\
 \\
  IBM T. J. Watson Research Center & Dept. of Physics \\
  P. O. Box 218 & University of California Los Angeles \\
  Yorktown Heights, NY~~10598	& Los Angeles, CA~~90024 \\
  divince@watson.ibm.com & smolin@vesta.physics.ucla.edu
\end{tabular}}

\maketitle

%I don't know why I have to reset thispagesyle, but otherwise get page numbers
\thispagestyle{empty}

\subsection*{\centering Abstract}
%IEEE allows italicized abstract
{\em
We present numerical results which show how two-bit logic gates can be
used in the design of a quantum computer.  We show that the Toffoli gate,
which is a universal gate for all classical reversible computation,
can be implemented using a particular sequence of exactly five two-bit
gates.  An arbitrary three-bit unitary gate, which can be used to build up any
arbitrary quantum computation, can be implemented exactly with six two-bit
gates.  The ease of implementation of any particular quantum operation
is dependent upon a very non-classical feature of the operation,
its exact quantum phase factor.
}

\section{Introduction}

The quantum computer as a desirable future technology, which if achievable
would revolutionize at least some areas of computational science, need hardly
be promoted by the present contribution; the reader need only glance at
some of the other papers presented at this conference\cite{Shor,Bert}
to see the great promise (and perhaps overenthusiasm\cite{Land})
of this field.

A quantum computer may be contrasted with a classical computer in the
following way:  In a classical computer the elementary operations are
boolean logic operations which are applied to a current state of the bits
of the computer, resulting in a new bit state after a short operation time.
These elementary logical operations are performed by {\em logic gates}, which
typically operate on pairs of input bits, producing one output bit (e.g.,
XOR, AND, etc.).  The quantum computer is very different:  its elementary
operations are unitary transformations applied to a wavefunction, that
wavefunction being expressed as a complex linear superposition of the
possible states of the computer, that is, of the possible states of its
bits.

As in traditional computer design, it is necessary that the complete
computation which is desired (i.e., the overall unitary transformation
representing the complete computation) should be broken down into a
discrete sequence of elementary logical operations.  Thus, the ``quantum
logic gate" executes a unitary transformation which operates only on the
part of the state of the system (i.e., the wavefunction) which involves
a small subset of the bits of the computer.
Such a ``gate" involves a physical interaction among only the
small subset of bits, and in this respect it is similar to a conventional
logic gate.  It differs from the classical logic gate in that the number
of output bits must equal the number of input bits; quantum processes are
reversible, and must involve no destruction of information.

Only a few useful quantum computations are known
presently\cite{Shor,Woot,Copp},
but we can expect that more may be discovered in the future.  Therefore,
it is legitimate to ask the following general question:  Given an arbitrary
desired computation $U(2^n)$ (i.e., arbitrary unitary operation on $n$ bits),
how may it be implemented by a sequence of quantum logic gates?  Deutsch
went most of the way towards answering this question in 1989\cite{Deut},
when he showed that just a single three-bit quantum logic gate is ``universal",
i.e., that a concatenation of this one gate applied in turn to different
triplets of bits, could implement any $U(2^n)$.  As an $8\times 8$ matrix
applied to the three-bit states $|0,0,0>$, $|0,0,1>$, ... $|1,1,1>$,
the S-matrix of Deutsch's
gate is:
\renewcommand{\arraystretch}{0.65}
\begin{equation}
{\bf U}_D=\ \left(\begin{array}{cccccccc}1&\ &\ &\ &\ &\ &\ &\ \\
\ &1 &\ &\ &\ &\ &\ &\ \\ \ &\ &1 &\ &\ &\ &\ &\ \\ \ &\ &\ &1 &\ &\ &\ &\ \\
\ &\ &\ &\ &1 &\ &\ &\ \\ \ &\ &\ &\ &\ &1 &\ &\ \\
\ &\ &\ &\ &\ &\ &i\cos\lambda &\sin\lambda \\
\ &\ &\ &\ &\ &\ &\sin\lambda &i\cos\lambda \end{array}\right)\label{four},
\end{equation}
where $\lambda$ is any constant such that $2\lambda/\pi$ is an irrational
number.

Deutsch's work left a few issues unsettled, however.  He did not
address the question of whether there might be a two-bit quantum logic gate
which is universal.  It was widely suspected that there was not, because
it was known that no set of two-bit gates was sufficient for another formal
system, classical reversible computation.
Toffoli\cite{Toff} showed that for the classical problem, the minimal universal
gate operates on three bits (this is now known as the Toffoli gate).
Nevertheless,
recently one of us showed
that Deutsch's three-bit gates could be created by a sequence of two-bit
quantum gates\cite{Divi}
(Deutsch's
work had never excluded this possibility).

Deutsch's work also had nothing to say about the {\em efficiency} of the
implementation of the $U(2^n)$ computation by quantum gates, and it is this
issue which we will address in the present paper.  Indeed, his calculation
\cite{Deut} only showed that an {\em infinite} sequence of $U(2^3)$ operations
would be capable of approximating, to arbitrarily high accuracy, a desired
$U(2^n)$; likewise, the recent proof of the universality of two-bit gates
only shows that an infinite sequence of $U(2^2)$ operations
would be capable of approximating a desired
$U(2^3)$.  A recent paper\cite{Reck} shows how any $U(N)$ can be rewritten {\em
exactly} as a finite sequence of simpler unitary operations; however,
the elementary operations used in that paper do not operate on a small
number of bits, and so do not qualify as ``gates" in the present sense
of the word.  (See also \cite{Eker} for a very closely related result.)

In this paper, then, we present a sequence of numerical experiments which
explore the issue of efficient implementation.  We show that
an arbitrary $U(2^3)$ can indeed be expressed {\em exactly} using a sequence
of (unconstrained) two-bit gates, and that the number of gates necessary
for this exact implementation is {\em six}.  We will also present the
corresponding results for three-bit gates of special importance.  We show
that the Toffoli gate also requires exactly five two-bit gates for
its exact implementation, which is somewhat surprising given its simple
form.  It has recently been shown\cite{Marg}
that Toffoli and Fredkin gates, but
with the wrong
quantum phases, can be exactly implemented in just {\em three} two-bit gates.

We will present a few other calculations illustrating more generally that
the requirement of exact phase control\cite{Denk} can substantially increase
the complexity of implementation.  Still, we view the bound of 6$\times$
on the
complexity of implementation using two-bit operations as a hopeful one;
it seems likely that it will often be desirable to pay this
6$\times$ complexity
cost to be able to implement more physically realizable
gates (see Conclusions).  We will {\em not} in this paper address the
larger implementation problem suggested by Deutsch's work, namely that
of efficiently implementing the
arbitrary $U(2^n)$ with a finite number of gates.
Hopefully the present investigation will provide a basis on which to
proceed on this harder problem.

\section{The Simulations}

We have attacked the problems listed above using ``brute force" numerical
experiments.  While this approach generally does not permit any result to
be rigorously proved, it does provide a quick way to generate a relatively
large number of almost-certainly-correct results in an automatic way.  While
purely numerical, our approach takes advantage of a set of analytic
results which could be more generally useful, so we mention them now.

\subsection{Notation and Constraints}

\begin{figure}
\psfig{file=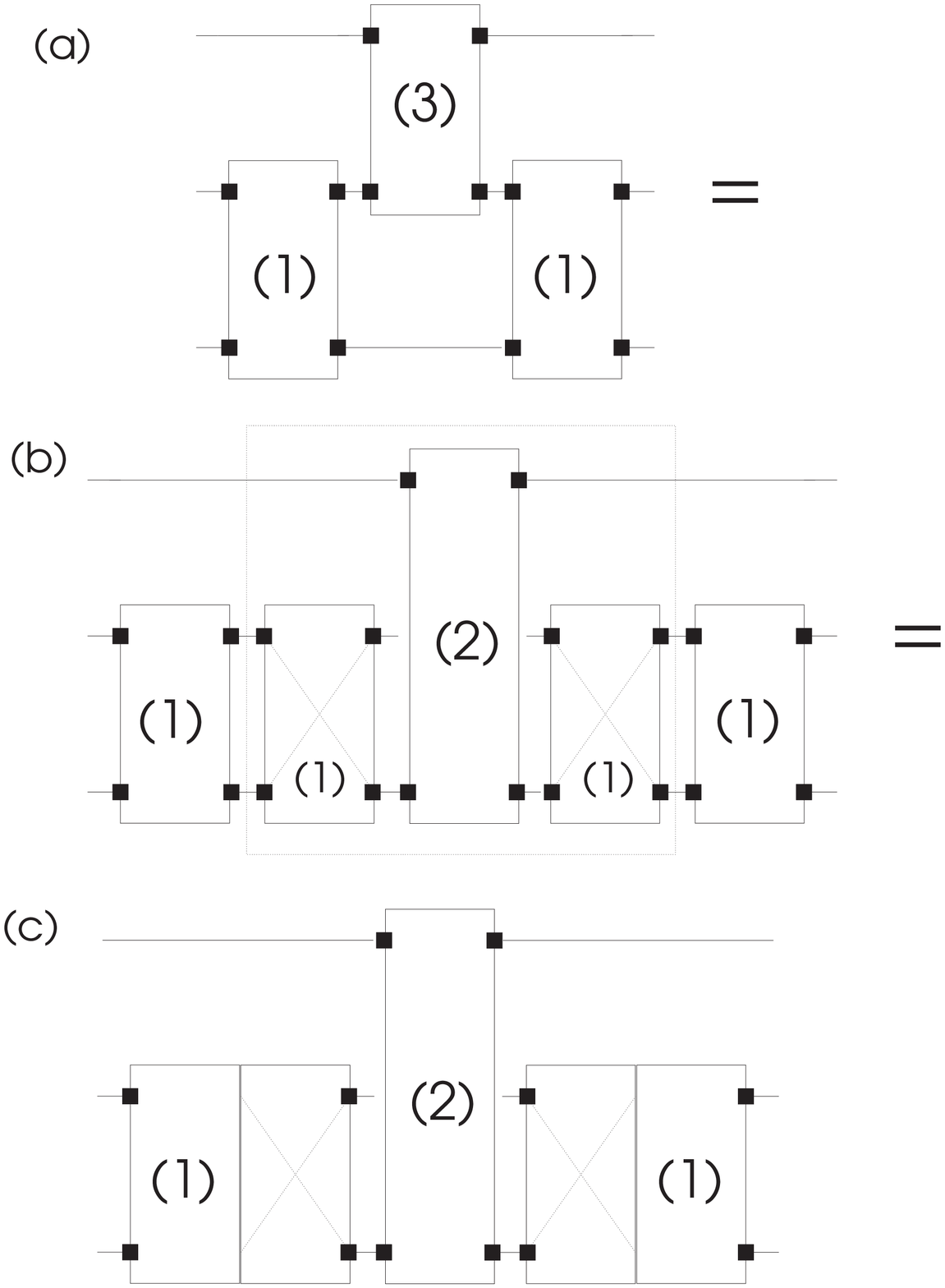,width=5in}
\caption{
Illustration of a replacement of a three-bit gate by a sequence of one-
and two-bit gates.  The numbering convention (1,2,3) for three bits is
indicated.  The one bit ``N" gate may be absorbed into the definition of
the two-bit gates.  According to the notation discussed in the text, the
network shown has the topology (1212).  For more details of the use of
this particular decomposition, see \protect\cite{Divi}.
\label{first}}
\end{figure}

Figure \ref{first} illustrates the kinds of constructions which we have
explored
with the computer.  It illustrates a particular
analytic equivalence discovered in
Ref. \cite{Divi}, between a particular three-bit gate designated
``$U_\lambda$" and a sequence of four two-bit gates (ignoring the two
one-bit gates labeled ``N" for the moment).  Ref. \cite{Divi} gives
more details of the calculations which this diagram depicts.
The horizontal lines
in the figure may be taken to denote the world lines of the three bits (they
may be spin-1/2 degrees of freedom in a physical implementation).  The
symbols are drawn to show that the first and third gates (the ``X"'s) operate
on bits 2 and 3 only, and the second and fourth gates (``V"'s) operate
on bits 1 and 3 only.

In the numerical work, we explore an exhaustive set of two-bit gate topologies
up to a certain maximum size.  To describe our classification
of these topologies, it is convenient to have a very compact notation
for a particular arrangement of gates.  We will use the following convention:
the function of a two-bit gate will be denoted by the number of the bit
on which it does {\em not} act.  Obviously this is not a good notation
for a general network, but suffices for the study of three-bit networks which
we conduct here.  Thus, by this notation, the four-gate sequence in
Fig. \ref{first}
is denoted (1212).  Note that this designation is unchanged by the presence
of the one-bit N gates; they can be absorbed into the X gates, with a
corresponding redefinition of their operation.  We will assume throughout
this section that an arbitrary two-bit operation is permissible at every
position.  We will discuss briefly in the Conclusions the physical feasibility
of this rule.

In the following computation it will be important to know all the possible
distinct gate topologies, especially those with four or five two-bit gates.
Two different topologies can be equivalent as
networks for the following reasons:

{\em Time-Reversal.}  When a sequence of unitary gates is operated in
time-reversed order, the inverse unitary operation results.  In all the
cases which we consider below, the desired unitary operation is either
self-inverse, or belongs to an ensemble which contains the inverse of
every matrix.  This means that the time-reversed version of any topology
is equally powerful in implementing the desired matrix or ensemble of matrices.
So, for example, topologies (12123) and (32121) are equivalent, and we will
write as shorthand (12123)=(32121).

{\em Bit Relabeling.}  The unitary matrices which we want to implement are
in many cases invariant with respect to relabeling of the bits.  If for
example, the matrices are invariant under a relabeling of bits 1 and 2,
then (12123)=(21213).

\begin{figure}
\psfig{file=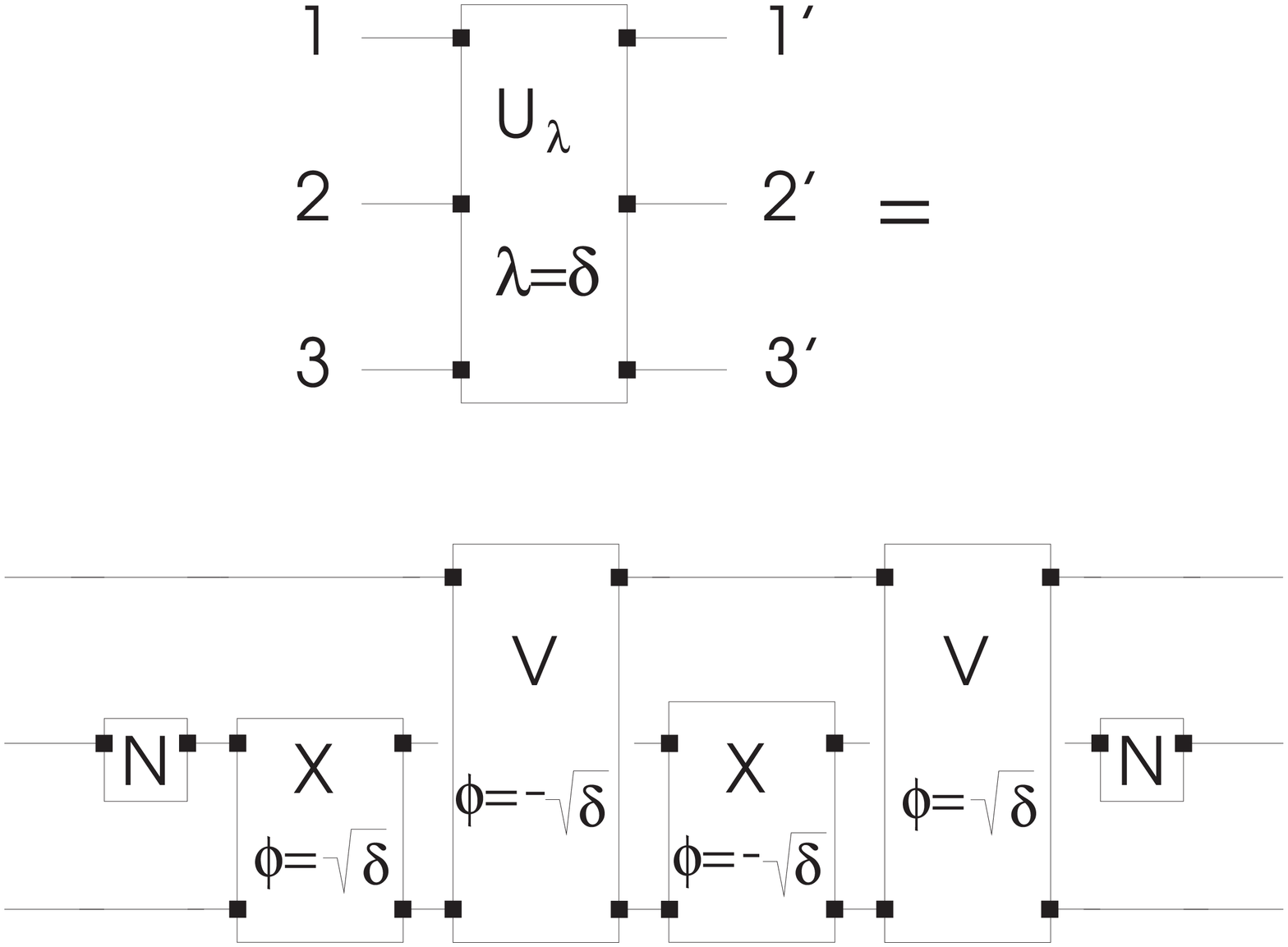,width=6in,angle=90}
\caption{
Modification of network topology by swapping.  The (131) topology of (a) is
changed in (b) by replacing (3) by the set of gates in the dotted box: a (2)
and two surrounding swapping gates of the (1) type.  The pairs of (1)'s are
then merged together in (c).  The net result is the replacement
(131)$\rightarrow$(121).
\label{second}}
\end{figure}

{\em Conjugation by Swapping.}  The swapping of the states of any pair of
bits is a special two-bit logic gate\cite{Deut},
and this gate can be used to change
the network topology.  Figure \ref{second}
shows how this is done; the (3) gate in
part a) can be changed to a (2) gate in part b) by the insertion of
(1)-type swap gates, that is, those which interchange the states of bits
2 and 3.  This operation, the replacement (3)$\rightarrow$(121),
is a conjugation in
the group-theoretic sense.  It is generally not useful in identifying
equivalent topologies of the same length, since it generally increases
the number of gates.  However, if a gate is surrounded by two gates
operating on the same bits, then the swap gates can absorbed into
the adjoining gates, as shown in Fig. \ref{second}(c).
Thus, the general equivalence
which conjugation implies is (iji)=(iki), where i, j, and k are all
different.  For the five-gate network used as an example above, it
produces the relations (12123)=(13123)=(12323)=(12313).

Below we will use these three operations to produce irreducible sets
of topologies to examine numerically.

\subsection{Numerical Technique}

To determine whether a particular three-bit operation can be implemented
using a given network of two-bit gates, we employ a non-linear minimization
procedure.  The objective function for this minimization is constructed as
follows:  For each gate in the network, an $8\times 8$
matrix is set up with a completely
{\em general} parameterization of the gate.  For example, if the two-bit
gate is of type (3), then its S-matrix has the form
\begin{equation}
S =\ \left(\begin{array}{cc}U(4)&0\\0&U(4)\end{array}\right)
\label{two},
\end{equation}
i.e., the matrix has a $4\times 4$ block-diagonal form, with two
identical copies
of an arbitrary U(4) matrix on the diagonal.  The type (1) and (2) matrices
are the same except for an appropriate re-ordering of the rows and columns.
$S$ has $4^2=16$ free real-valued parameters\cite{Math}.

Now, the unitary matrix of the complete network is given by a product
of S matrices as in Eq. \ref{two}; call the resulting matrix $S^{tot}$.
If the number of gates in the network is $N_2$, $S^{tot}$ has $16N_2$
free parameters (not necessarily all independent).  If the desired three-bit
operation has S-matrix $U$ (we will refer to this as the ``target matrix"
below),
then we form the following objective function
of $16N_2$ variables:
\begin{equation}
f = \sum_{i=1}^{8}\sum_{j=1}^{8}|U_{ij}-S^{tot}_{ij}|^2.\label{three}
\end{equation}
We then seek minima of $f$ numerically.  If we find a minimum at which
$f=0$ to reasonable numerical accuracy, then we have found the desired
two-bit implementation of the three-bit gate.  If all of the minima of
the function have $f>0$, we conclude that the desired three-bit operation
cannot be implemented by the two-bit topology being considered.

The numerical routine used was a variable-metric
minimizer from the Harwell subroutine library, specifically the VF04AD
routine.  This routine introduces the unitarity of the individual
matrices using Lagrange multipliers, and Broyden-Fletcher-Goldfarb-Shanno
minimization is used\cite{Numr};
the user is not required to supply expicitly a matrix of derivitives,
which is helpful in dealing with complicated nonlinear functions
such as those we have here.  This routine is, of course, only capable
of finding local minima, having no way to determine if the minimum
is indeed global.  Many such local minima are found in the present
calculations,
so the minimizer
was in each case run starting from several sets of random initial conditions
to ensure the real minimum could be found.

\subsection{Results: arbitrary U(8)}

We have obtained convincing evidence that any U(8) target gate may be obtained
by a network of six two-bit gates, and no fewer.  We demonstrate this as
follows:  First, we find that using the six-gate topology (121212),
a minimum of the objective function is obtained which is essentially zero,
{\em viz.}, $f_{min}\leq 10^{-5}$.  We have confirmed this with three different
U(8) matrices generated randomly.  (They are actually constructed from the
eigenvectors of a random Hermitian matrix.)  Second, we perform an exhaustive
study of all distinct five-gate topologies.  Using the symmetries mentioned
above which the ensemble of all U(8) matrices possess--invariance under
inversion, and under all relabelings of bits 1, 2, and 3--we need only
consider two distinct five-gate topologies, (12121) and (12123).
For neither of these topologies do we find a solution; in every case,
the minimum value of the objective function $f$ is $f_{min}\geq 2.0$.

This shows, of course, that no topologies with fewer than 5 gates can implement
the arbitrary U(8).  It is worth noting, though, that
there are simple {\em analytic} arguments that an arbitrary U(8) cannot be
implemented in fewer than five gates.  The number of parameters in the
general U(8) is $8^2=64$\cite{Math}.  The number of distinct parameters
in the objective function for a network with $N_2$ gates is $16N_2-(N_2-1)$.
The second term in this expression comes from absorbing the overall phase
factor of each two-bit gate into a single global phase factor (this amounts
to considering SU(4) rather than U(4) matrices).  Simple arithmetic shows
that $N_2$ must be at least 5 in order that the objective function have
at least 64 free parameters.

\subsection{Results: Toffoli gate}

In all known examples of quantum computing\cite{Shor,Bert,Woot} it is
essential that classical reversible logic gates be executed as quantum
operations.  All classical reversible logic can be generated by the
Toffoli gate\cite{Toff},
\begin{equation}
{\bf U}_T=\ \left(\begin{array}{cccccccc}1&\ &\ &\ &\ &\ &\ &\ \\
\ &1 &\ &\ &\ &\ &\ &\ \\ \ &\ &1 &\ &\ &\ &\ &\ \\ \ &\ &\ &1 &\ &\ &\ &\ \\
\ &\ &\ &\ &1 &\ &\ &\ \\ \ &\ &\ &\ &\ &1 &\ &\ \\
\ &\ &\ &\ &\ &\ &0 &1 \\
\ &\ &\ &\ &\ &\ &1 &0 \end{array}\right)\label{five},
\end{equation}
so we have explored the minimal
implementation of this U(8) element.  Of course, from
the preceding section we know that it is implementable with six gates
in the (121212) topology.
We also find that the Toffoli gate can be achieved with five gates.  We
find that certain topologies work ((12123), (12132), (12312), (13213)) and
others do not ((12121), (31213))\cite{alll}.  However, no simpler
implementation of $U_T$ is possible, which we established by finding
that the minimum of the objective function Eq. (\ref{three}) was of order
unity for the four-gate topologies (1231) and (3123).  Thus, by the
criterion of length of its two-gate implementation, the Toffoli gate
is only slightly less complex that the general three-bit gate!

Previously, the best known implementation of the Toffoli gate was a
six-gate implementation of topology (121231) found by Coppersmith\cite{Copp2}.
(We do not obtain the same matrices as Coppersmith
when we study this topology numerically, which is indicative of a weakness
of our approach: since there are many unneeded free parameters in the
(121231) implementation, our procedure does not pick out the
one with ``simple" two-bit gates which Coppersmith found.)
Sleator and Weinfurter\cite{Slea} have also recently shown a two-bit-gate
implementation of $U_T$; while their implementation involves a greater
number of gates, it respects the practical constraints involved in
implementing unitary transformations in cavity QED, which
of course have also not been considered here.

\subsection{Results: Margolus' modification of Toffoli--the phase problem}

Our results demonstrate that the ease of implementation of the quantum gate
depends very markedly upon its phases, a situation which has no classical
analog.  To illustrate this, we quote a result from N. Margolus\cite{Marg}
for the implementation of a slight modification of the Toffoli gate:
\begin{equation}
{\bf U}_M=\ \left(\begin{array}{cccccccc}1&\ &\ &\ &\ &\ &\ &\ \\
\ &1 &\ &\ &\ &\ &\ &\ \\ \ &\ &1 &\ &\ &\ &\ &\ \\ \ &\ &\ &1 &\ &\ &\ &\ \\
\ &\ &\ &\ &1 &\ &\ &\ \\ \ &\ &\ &\ &\ &-1 &\ &\ \\
\ &\ &\ &\ &\ &\ &0 &1 \\
\ &\ &\ &\ &\ &\ &1 &0 \end{array}\right)\label{six},
\end{equation}
which differs from $U_T$ only in that it changes the phase of the $|1,0,1>$
component of the quantum state.  Despite this minor change, Margolus has shown
that $U_T$ can be implemented with just {\em three}
two-bit gates, in the (121) topology.  (By checking (12), (13) and (23),
we have shown that Margolus' result is minimal.  We also find that for
the (121) topology, other choices of phase
factors in the matrix Eq. (\ref{six}) besides Margolus'
can be implemented.)

To further illustrate the difficulty of achieving the desired phase in quantum
gates, we investigated the two-bit implementation of perhaps the simplest
possible three-bit gate, which Deutsch\cite{Deut} calls ``X":
\begin{equation}
{\bf U}_\phi=\ \left(\begin{array}{cccccccc}1&\ &\ &\ &\ &\ &\ &\ \\
\ &1 &\ &\ &\ &\ &\ &\ \\ \ &\ &1 &\ &\ &\ &\ &\ \\ \ &\ &\ &1 &\ &\ &\ &\ \\
\ &\ &\ &\ &1 &\ &\ &\ \\ \ &\ &\ &\ &\ &1 &\ &\ \\
\ &\ &\ &\ &\ &\ &1 &\ \\
\ &\ &\ &\ &\ &\ &\ &e^{i\phi} \end{array}\right)\label{seven},
\end{equation}
for which the state is unchanged except for the modification of one phase
factor.  It turns out that even this simple matrix is not very simple
to implement with two-bit gates.
It can of course be constructed in the ``universal" 6-gate topology
(121212); the 5-gate topologies (12123) and (12312) work, although (12121)
fails; and, no 4-gate topology can implement $U_\phi$, as we established by
checking (1231) ((1212) is already excluded by the 5-gate result).  So,
its complexity is the same as $U_T$'s.
As Denker has recently emphasized\cite{Denk}, having incorrect phases
in a quantum computation could be catastrophic, e.g., for a quantum Fourier
transform operation\cite{Shor,Copp}; thus, $U_M$ can be used in place of
$U_T$ only with great care, and ``simple" phase adjustments like $U_\phi$
must be carefully designed.  The present results show that these seemingly
minor adjustments may exact a considerable cost in network complexity.

\section{Concluding Remarks}

There are several obvious weaknesses to the present results, which we would
like to touch on.  First, since the approach is purely numerical, no result
in this paper can be taken as strictly proved, only strongly suggested.  In
\cite{Divi} we proved, using the generator calculus of the Lie groups, that
representation of any three-bit gate by a
sequence of two-bit gates was possible,
provided that arbitrarily many two-bit gates are permitted.  To our
disappointment, we have found no way of adapting this calculus to prove
the stronger results, that operations are implementable {\em exactly} in a
{\em finite} number of gates.  Hopefully by producing a set of definite
results here, we may be led to some new way of applying Lie group theory
to obtain the desired stronger analytic results.

A second weakness of the present approach is a more physical one:  we assume
that {\em any arbitrary} two-bit unitary operation is possible (see Eq.
(\ref{two})).  Whether this is true or not depends on the details of the
physical implementation of the quantum system.  As Ref. \cite{Slea}
illustrates, for any specific system (e.g., cavity QED) certain two-bit
operations are much {\em easier} to implement that others.  It seems likely
that in most circumstances where several different two-bit operations are
easy to accomplish, that suitable concatenations of these operations can
generate any arbitrary $U(4)$; probably the most is known about this in
the context of multiple-resonance NMR\cite{Slic}.  Still, no definite
results have been obtained on this question yet, and it should be the
subject of future study.  It would also be sensible to include the ease
of construction of particular classes of $U(4)$ operations into our
numerical optimizations.

To summarize: Our numerical studies indicate that any arbitrary three-bit
quantum gate can be produced exactly as a concatenation of six
two-bit quantum gates.  A particular three-bit logic gate of great importance,
the Toffoli gate, can be produced with five.  Relaxing the constraint that
the Toffoli gate introduce no phase shifts, which is often a dangerous thing
to do, leads to a three-gate implementation with two-bit operations, as
shown by Margolus.  The insertion of even the simplest additional phase
shift in the three-bit space requires five additional two-bit gates.

%this is how to do an unnumbered subsection
\subsection*{Acknowledgements}
We are grateful to C. H. Bennett, H. J. Bernstein and R. Landauer for
many helpful discussions.

\vfill
\pagebreak
{}~~

\clearpage

\clearpage

\begin{thebibliography}{99}

\bibitem[*]{byline}
Also at IBM Thomas J. Watson Research Center.

\bibitem{Shor}
P. W. Shor, ``Algorithms for quantum computation:
discrete log and factoring", these proceedings; see also
Proceedings of the 35th IEEE Symposium on the Foundations of Computer Science,
1994 (to be published).

\bibitem{Bert}
A. Berthiaume, D. Deutsch, and R. Jozsa, ``The stabilisation of quantum
computations", these proceedings.

\bibitem{Land} R. Landauer, ``Zig-zag path to understanding", these
proceedings; see also R. Landauer, ``Is quantum mechanics useful?", Proc.
Roy. Soc. Lond., (to be published).

\bibitem{Woot} W. K. Wootters, ``Quantum block coding", unpublished (1994).

\bibitem{Copp} D. Coppersmith, ``An approximate Fourier
transform useful in quantum factoring",
IBM Research Report RC19642 (1994).

\bibitem{Deut} D. Deutsch, ``Quantum computational networks",
Proc. Roy. Soc. Lond. A {\bf 425}, 73 (1989).

\bibitem{Toff} T. Toffoli ``Reversible Computing", in {\em Automata,
Languages and Programming}, eds. J. W. de Bakker and J. van Leeuwen (Springer,
New York, 1980), p. 632; Technical Memo MIT/LCS/TM-151, MIT Lab. for
Comp. Sci. (unpublished).

\bibitem{Divi} D. P. DiVincenzo, ``Two-bit gates are universal for quantum
computation", submitted to Phys. Rev. A (1994).

\bibitem{Reck} M. Reck, A. Zeilinger, H. J. Bernstein, and P. Bertani,
``Experimental realization of any discrete unitary operator", Phys. Rev.
Lett. {\bf 73}, 58 (1994).

\bibitem{Eker} A. Ekert and R. Jozsa, ``Notes on Shor's efficient algorithm
for factoring on a quantum computer", Workshop on
Quantum Computing and Communication, Gaithersburg, MD, August 18-19, 1994,
to appear on WWW.

\bibitem{Marg} N. Margolus, private communication.
%%A0794 NOTEBOOK, 7/11/94, "quantum computation paper"

\bibitem{Denk} A point recently anticipated by J. S. Denker, Workshop on
Quantum Computing and Communication, Gaithersburg, MD, August 18-19, 1994
(unpublished).

\bibitem{Numr} W. H. Press, B. P. Flannery, S. A. Teukolsky, W. T.
Vetterling, {\em Numerical Recipes} (Cambridge University Press, 1986),
Chap. 10.

\bibitem{Math} J. Mathews and R. L. Walker, {\em Mathematical
Methods of Physics}, (Benjamin, Second Edition, 1970), Chap. 16, has a
basic review of the theory of unitary matrices.

\bibitem{alll} These topologies are all the distinct ones, taking account
of the fact that the Toffoli gate is not invariant under any bit relabeling
involving bit 3.

\bibitem{Copp2} D. Coppersmith, unpublished (1994).

\bibitem{Slea} T. Sleator and H. Weinfurter, ``Quantum teleportation and
quantum computation based on cavity QED", Ann. NY Acad. Sci., to be published.
(Presented at the Conference on Fundamental Problems in Quantum Theory, June
18-22, 1994, Baltimore, MD.)

\bibitem{Slic} C. P. Slichter, {\em Principles of Magnetic Resonance}
(Third Edition, Springer-Verlag, 1992).

\end{thebibliography}
\end{document}